\def\ANON{0}

\documentclass[sigconf]{acmart}
\usepackage{amsmath}
\usepackage{url}
\usepackage{hyperref}
\newtheorem{fact}{Fact}

\copyrightyear{2022} 
\acmYear{2022} 
\setcopyright{acmcopyright}\acmConference[ARES 2022]{The 17th International Conference on Availability, Reliability and Security}{August 23--26, 2022}{Vienna, Austria}
\acmBooktitle{The 17th International Conference on Availability, Reliability and Security (ARES 2022), August 23--26, 2022, Vienna, Austria}
\acmPrice{15.00}
\acmDOI{10.1145/3538969.3544432}
\acmISBN{978-1-4503-9670-7/22/08}
\begin{document}
\title{Revisiting a Privacy-Preserving Location-based Service Protocol
using Edge Computing}

%

\ifnum\ANON=0
\author{Santosh Kumar Upadhyaya}
\orcid{0000-0002-6829-1727}
\email{santosh.upadhyaya@iiitb.ac.in}
\affiliation{
   \institution{IIIT, Bangalore}
   \country{IN}}

\author{Srinivas Vivek}
\orcid{0000-0002-8426-0859}
\email{srinivas.vivek@iiitb.ac.in}
\affiliation{
   \institution{IIIT, Bangalore}
   \country{IN}}
\else
\author{}
\affiliation{}
\fi


\begin{abstract}
 Location-based services are getting more popular day by day. Finding nearby 
 stores, proximity-based marketing, on-road service assistance, etc., are 
 some of the services that use location-based services. In location-based 
 services, user information like user identity, user query, and location must 
 be protected. Ma et al. (INFOCOM-BigSecurity 2019) proposed a privacy-preserving 
 location-based service using Somewhat Homomorphic Encryption (SHE). Their
 protocol uses edge nodes that compute on SHE encrypted location 
 data and determines the $k$-nearest points of interest contained in
 the Location-based Server (LBS) without revealing the original user 
 coordinates to LBS, hence, ensuring privacy of users locations.

In this work, we show that the above protocol by Ma et al. has a critical
flaw. In particular, we show that their secure comparison protocol has a correctness
issue in that it will not lead to correct comparison. A major consequence 
of this flaw 
is that straightforward approaches to fix this issue will 
make their protocol insecure. Namely, the LBS will be able to recover the
actual locations of the users in each and every query.

\end{abstract}



\maketitle              

\section{Introduction}
Location-based service is meant to provide real-time information based on the current location of a user by combining multiple entities like the global positioning systems, information and communication systems, and the Internet \cite{schiller2004location, junglas2008location}. User identity, location and query information are sensitive and personal, and, hence, must be protected \cite{shin2012privacy, 8360466}. This information needs to be protected as this could potentially be misused \cite {8316781}. Hence, it is in the interest of a Location-based Service (LBS) provider to protect the private information of the users to maintain its reputation, and, hence, its business itself. 
Ma et al. proposed a privacy-preserving location-based service using Somewhat Homomorphic Encryption (SHE) \cite{ma-infocom19}. The user provides his/her
encrypted location information and the encrypted query to an Edge Node (EN).
The location coordinates are encrypted using an SHE, while the query is
encrypted using a traditional encryption scheme.
When the encrypted service request and the encrypted user location coordinates reach EN, it generates an encrypted virtual location using a standard K-anonymity technique, in turn 
referring to the historical location information \cite{8274909}.The LBS server contains the location coordinates of many points of interest. This information is not encrypted. Depending on the user's query, it will select a subset
of these points. But this selection must be done using the encrypted 
coordinates of the user virtual location computed by EN and the points of interest stored as plaintexts in LBS. The main 
purpose is to securely choose, say, $k$, nearest points of interest around 
the user's location \cite{8560131}. The metric used here is the Euclidean distance. So the 
crux of the protocol of Ma et al. is an efficient privacy-preserving distance comparison
protocol that is executed between EN and LBS. The detailed steps are recalled 
in the Section \ref{sec:recap}.



\noindent\textbf{Our Contribution.} We show, in Section \ref{sec:analysis}, that the privacy-preser\-ving 
distance comparison protocol of Ma et al., that is eventually used in determining 
the nearest points of interest, suffers from a correctness flaw. Namely, the 
output of this comparison protocol is \textit{not} necessarily correct. 
A major consequence of this flaw is that a straightforward approach to fix this
flaw would be to give out the LBS the (signed) differences of the distances. 
We show, for the sake of completeness, that using these differences an LBS will be able to
recover the actual location coordinates in each and every user query. We also 
consider another straightforward modification of the protocol whereby the 
differences of distances are masked by an independently chosen random value but that
still allows for efficient comparison. We again show
that this approach too fails in preserving the privacy of the user locations.
Our work demonstrates that fixing the protocol of Ma et al. is non-trivial
without incurring a significant cost. 

\section {Recap of the Protocol from Ma et al.}
\label{sec:recap}
In this section, we briefly recollect the steps of the protocol from  \cite{ma-infocom19}.
There are four different entities in the protocol:
\begin{itemize}
\item User
\item Edge Node (EN)
\item Location-based Services (LBS) Server
\item Certificate Authority (CA)
\end{itemize}

\begin{figure*}[ht]
  \centering
  \includegraphics[width=\linewidth]{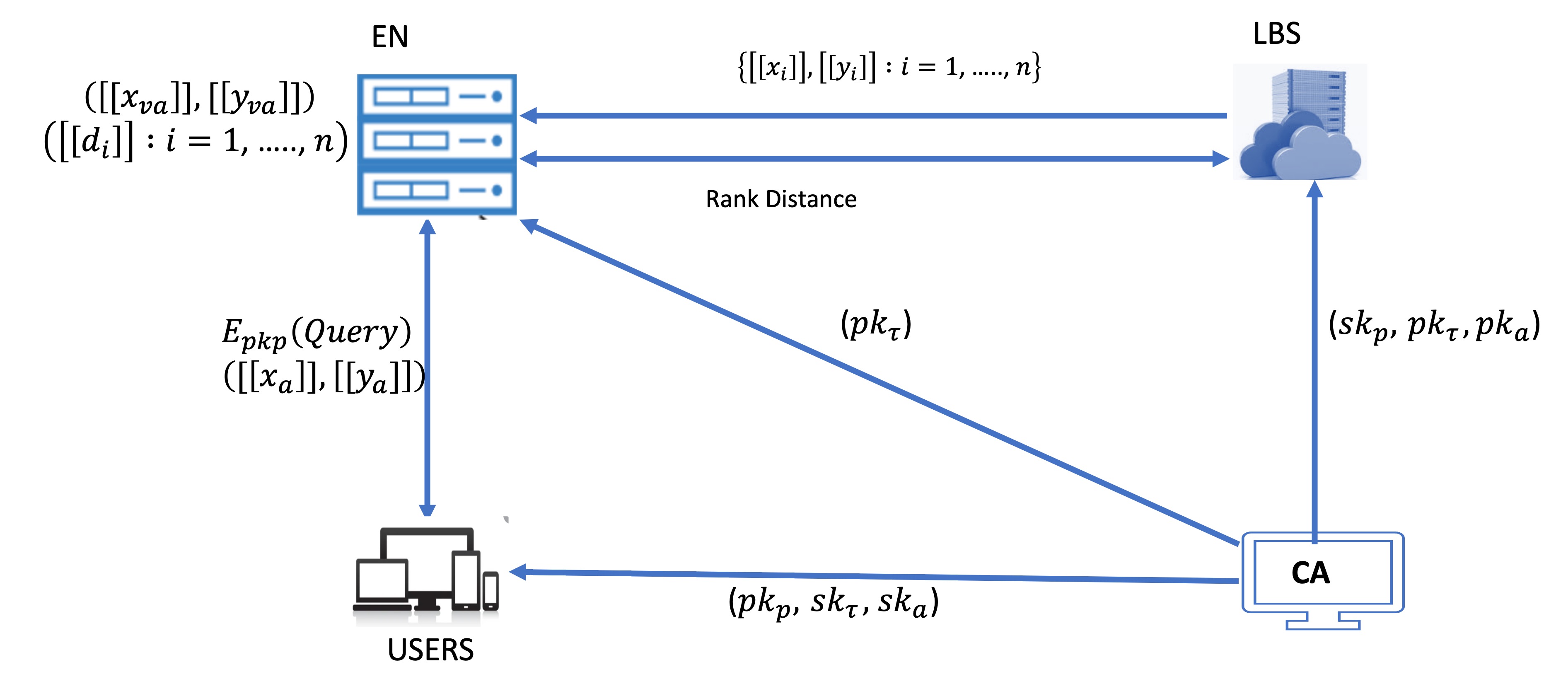}
  \caption{Entities involved in the protocol from \cite{ma-infocom19}.}
\end{figure*}



\subsection{Initialization}
During the initialization step, the user registration process is executed. When a user requests for a location service, CA sets up the required private and public keys for the user, EN, and LBS. The CA generates 3 pairs of public and private keys.
\begin{itemize}
\item $pk_P, sk_P$:
$pk_P$ is sent to the user to encrypt the user request. $sk_P$ is sent to the LBS to decrypt the user query.
\item $pk_\tau$ is sent to the LBS to encrypt the query response using SHE \cite{DBLP:journals/iacr/FanV12}.
$sk_\tau$ is sent to the user to decrypt the same.
\item $pk_\tau$ is sent to the EN which is used to calculate the encryption of $z$.
\item $pk_a$, $sk_a$:
$pk_a$ is sent to the LBS and $sk_a$ is sent to the user.
\end{itemize}
Note that in this protocol the FV SHE scheme is used to encrypt users' location information. Recall that 
if $m$ and $m'$ are plaintexts and their corresponding ciphertexts are $c$ and $c'$, then  
using an SHE scheme, the encryption of $m + m'$ or $m \cdot m'$ can be derived from $c$ and $c'$ without the need to decrypt the ciphertexts. If $\llbracket x 
\rrbracket$ is a SHE ciphertext of a plaintext $x$, then $\llbracket m + m' \rrbracket$ can be computed as $\llbracket m \rrbracket + \llbracket m' \rrbracket$.
Similarly,  $\llbracket m \cdot m' \rrbracket$ can be computed as $\llbracket m \rrbracket \cdot \llbracket m' \rrbracket$.

\subsection{User Query: USER to EN}
To preserve the privacy of the user location, a K-anonymity based technique is used \cite{6831391}. When the user creates a query, the query is encrypted with $pk_P$ and the encrypted user query is sent to the nearest EN. The user sends the SHE encrypted location coordinates $\llbracket X_{a} \rrbracket$ and $\llbracket Y_{a} \rrbracket$ to EN. The EN is equipped with better storage and computing power in comparison to the user device \cite{8336969}. The EN only knows that the user is within its coverage area. The EN calculates the virtual address of the user by fetching the historical location information. If the current user is considered as the $\eta$th user, then EN fetches the (encrypted) location coordinates about the previous $t-1$ users:

\begin{equation} \label{eq:xva}
X_{va} = \frac{1}{t} (\llbracket X_{a} \rrbracket + \llbracket X_{\eta-1} \rrbracket + \llbracket X_{\eta-2} \rrbracket + \dots + \llbracket X_{\eta-t+1} \rrbracket)\\
\end{equation}

\begin{equation} \label{eq:yva}
Y_{va} = \frac{1}{t} (\llbracket Y_{a} \rrbracket + \llbracket Y_{\eta-1} \rrbracket + \llbracket Y_{\eta-2} \rrbracket + ... + \llbracket Y_{\eta-t+1} \rrbracket)
\end{equation}
$(X_{va}$,  $Y_{va})$ is the computed virtual location of the user. This is the classical \textit{moving average} technique used in statistics \cite{wiki:Moving_average}. $(\llbracket X_{va} \rrbracket$, $\llbracket Y_{va} \rrbracket)$ denotes its SHE ciphertext. The question that arises is how to compute the encryption of $\frac{1}{t}$. One could possibly use the fixed-point encoding scheme from \cite{CostacheSVW16} for this purpose. 

\subsection{EN to LBS}
The EN relays the encrypted user query to LBS. After receiving the user query from EN, LBS decrypts it using $sk_P$ and obtains the 
user query as plaintext. The LSB does not have any information about the virtual address $(\llbracket X_{va} \rrbracket, \llbracket Y_{va} \rrbracket)$ 
as it is encrypted. It needs to further interact with EN to build the query response. The EN’s territorial information is available 
with LBS. It fetches location coordinates of the services whose information it has, and sends them to EN after encrypting these location coordinates using the SHE scheme. If there are $n$ 
services supported by EN, then let the location coordinates for these $n$ services be $\{( x_{i} ,  y_{i} ): i = 1, ... n \}$. The encryption of these coordinates using key $pk_a$ are $\{ (\llbracket x_{i} \rrbracket, \llbracket y_{i} \rrbracket): i = 1, \ldots, n$ \}.

\subsection{LBS to EN} \label{LBS-EN-1}
The LBS sends $\{ (\llbracket x_{i} \rrbracket, \llbracket y_{i} \rrbracket): i = 1, ... n$ \} to EN. Once EN receives the information from LBS, it calculates $\{\llbracket d_{i} \rrbracket : i = 1, \dots, n \}$, which are the squared Euclidean distances of the
user virtual location to the $n$ services, as
\begin{equation} \label{eq:3}
\llbracket d_{i} \rrbracket = (\llbracket X_{va} \rrbracket -  \llbracket x_{i} \rrbracket) \cdot (\llbracket X_{va} \rrbracket -  \llbracket x_{i} \rrbracket) + (\llbracket Y_{va} \rrbracket -  \llbracket y_{i}\rrbracket)  \cdot (\llbracket Y_{va} \rrbracket - \llbracket y_{i} \rrbracket).
\end{equation}
In order to ensure the privacy of users locations, $\llbracket X_{va} \rrbracket$ and $\llbracket 
Y_{va} \rrbracket$ should not be sent to LBS, and so is the case with $\{\llbracket d_{i} \rrbracket : i = 1, \dots, n \}$.  Next, the LBS must somehow securely sort the encrypted (squared Euclidean) distances $\{\llbracket d_{i} \rrbracket : i = 1, \dots, n \}$ to determine
the nearest distance(s). An obvious way of sorting is to compare every pair of (encrypted) distances, and this is what is done next.
From the list of $\{\llbracket d_{i} \rrbracket : i = 1,..., n \}$, pick any two elements, say, $\llbracket d_{a} \rrbracket$ and $ \llbracket d_{b} \rrbracket$. 
Let $m$ be the maximum distance covered by EN. If the range is considered as a circular area, then $m$ is the diameter of the circle. In this case, $0 \leq d_{a}, d_{b} \leq m $. EN selects a number $l$ such that  
\begin{equation} \label{eq:4}
2 ^ l \geq m. 
\end{equation} 
The EN computes
\begin{equation} \label{eq:5}
\llbracket z \rrbracket = \llbracket 2 ^ l + d_{a} - d_{b} \rrbracket = \llbracket 2 ^ l \rrbracket + \llbracket d_{a} \rrbracket + \llbracket -d_{b} \rrbracket .\\
\end{equation}
$\llbracket z \rrbracket$ is the SHE ciphertext of $z$, and $z$ is an $l+1$-bit integer whose Most Significant Bit (MSB), $z_l$, depends on the value of $d_{a}$ and $d_{b}$. If the MSB of $z$ is 0, then $d_{a} < d_{b}$. Otherwise, $d_{a} \geq  d_{b}$. In order to indirectly send  the value $\llbracket z \rrbracket$ to LBS, EN creates a uniform random number $\rho$ of size $k + l + 1$ bits. Here, $k$ is the security parameter. The sum of $\llbracket z \rrbracket$ and $\llbracket \rho \rrbracket$ is computed, and is then sent to LBS:
\begin{equation} \label{eq:w}
\llbracket w \rrbracket = \llbracket z \rrbracket + \llbracket \rho \rrbracket .
\end{equation}

\subsection{EN to LBS}
Once LBS receives $\llbracket w \rrbracket$, it decrypts $\llbracket w \rrbracket$ and obtains $w$.
From $w$, it calculates $\bar w$ as follows: 
\begin{equation} \label{eq:wbar}
\bar w  = w \pmod{2 ^ l},
\end{equation} 
and then computes its SHE ciphertext $\llbracket \bar w \rrbracket$. 

\subsection{LBS to EN}
\label{subsec:wrho}
Let
\begin{equation} 
\label{eq:rhobar}
\bar \rho  = \rho \pmod{2 ^ l}.
\end{equation} 
Note that
$\bar w$ is available with LBS, and $\bar \rho$ is available with EN. Let ${(\bar w_{t-1}, \dots , \bar w{0})}$ and ${(\bar \rho_{t-1}}, \dots, \bar \rho_{0})$ be the bits of $\bar w$ and $\bar \rho$, respectively. The LBS encrypts each bit of $\bar w$ and is sent to EN.
It is proposed to compare $\bar w$ and $\bar \rho$ and determine the MSB of $z$, using which we can in turn compare $d_{a}$ and $d_{b}$. The DGK scheme 
\cite{DamgardGK07,DamgardGK09} is used for this  step.
LBS server then runs the DGK key generation algorithm to generate the public and private key pair.
The public key is sent to EN.
The below steps are run multiple times so that eventually the LBS learns the sorted order of $(d_{i} : i = 1, ..., n)$ . After this, the LBS builds the response to the user query and then sends it to the user. 

During this process, the LBS server chooses a random number between 1 and $-1$, and assigns it to $\epsilon$.
It calculates, for $j = 0, \ldots, l-1$,
\begin{equation} \label{eq:8}
\llbracket c_{j} \rrbracket = \llbracket \bar w_{j} \rrbracket \; \cdot \; \llbracket - \bar \rho_{j} \rrbracket \;\cdot\; \llbracket \epsilon \rrbracket \; \cdot\; \left(\prod_{v=j+1}^{l-1} \llbracket \bar w_{v} \oplus \bar \rho_{v}\rrbracket\right)^3
\end{equation}
EN randomly selects $\xi_j \in Z_{n}\; (j = 0, \dots l-1)$, and then it computes
\begin{equation} \label{eq:9}
\llbracket \bar c_{j} \rrbracket = \llbracket c_{j} \cdot \xi_{j} \rrbracket = \llbracket c_{j} \rrbracket ^ {\xi_{j}}. \end{equation}\\
Finally, EN would have $(\llbracket \bar c_{l-1} \rrbracket, \ldots, \llbracket \bar c_{0} \rrbracket)$.

\subsection{EN to LBS}
\label{subsec:last}
EN sends $(\llbracket \bar c_{l-1} \rrbracket, \dots, \llbracket \bar c_{0} \rrbracket)$ to LBS. After LBS receives this information, it decrypts this and checks the presence of 0. If 0 is present, that, the authors claim, indicates $\bar w > \bar \rho$, otherwise, $\bar w \leq \bar \rho$

The EN and LBS need to run these steps $n(n - 1)/2$ times. Finally, LBS will obtain the sorted order of $(d_{i} : i = 1, ..., n)$ without knowing anything about the values $(d_{i} : i = 1, ..., n)$. After the service locations to be sent is securely determined, the user query response is created and sent to EN after encryption with key $pk_\tau$. Finally, EN relays this query response to the user, and then the user decrypts it using its secret key $sk_\tau$.


\section {Correctness Flaw and Security implications on the Protocol of Ma et al.}
\label{sec:analysis}
In this section, we point out a critical flaw in the protocol from \cite{ma-infocom19} recalled in the previous section. 
This flaw corresponds to the steps of the protocol described in Sections \ref{subsec:wrho} and \ref{subsec:last}. Recall
that the idea behind these steps is to use $\bar \rho$ and $\bar w$ to determine $z_l$, the MSB of $z$ 
(see Equations \eqref{eq:w}, \eqref{eq:wbar} \eqref{eq:rhobar}). Recall that this bit $z_l$
is used to compare distances $d_{a}$ and $d_{b}$. The following is an elementary fact from arithmetic:
\begin{fact}
$z_l$ is independent of $\bar w$ and $\bar \rho$.
\end{fact}
This is because $\bar w$ is determined only by $\bar \rho$ and $\bar z$, and the latter is completely independent of $z_l$.
\begin{corollary}
The comparison protocol from \cite{ma-infocom19} does not correctly determine the comparison between encrypted distances.
\end{corollary}

The following toy examples illustrate the above observation.
\noindent\textbf{Example 1}

$z = 3  = (11)_{2}$

$l = 2 = (10)_{2}$

$\rho = 31 = (11111)_{2}$

$w = z + \rho = 34 = (100010)_{2}$


$\bar \rho = 31 \pmod{4} = 3 = (11)_{2}$

$\bar w = w \pmod {2 ^ 2} = 34 \pmod{4} = 2 = (10)_{2}$

\noindent\textbf{Example 2}

$z = 7  = (111)_{2}$

$l = 2 = (10)_{2}$

$\rho = 31 = (11111)_{2}$

$w = z + \rho = 38 = (100110)_{2}$


$\bar \rho = 31 \pmod{4} = 3 = (11)_{2}$

$\bar w = w \pmod {2 ^ 2} = 38 \pmod{4} = 2 = (10)_{2}$

In both the examples, the values of $\bar w$ and  $\bar \rho$ remain the same, but $z_l$ takes both $0$ and $1$. 

\subsection {Security Implications}

The privacy-preserving comparison protocol discussed above was proposed in \cite{ma-infocom19} in order to leak to LBS only
$z_l$, i.e., the result of comparison between any pair of distances. This was done because leaking the full value of $z$ would 
enable an adversary to determine the original user locations. For completeness, we briefly recollect next the steps to recover
$(X_a,Y_a)$, the original user location coordinates, when LBS obtains $\llbracket z \rrbracket$. Note that since the secure comparison protocol is flawed, giving 
out $z$ is a straightforward, but insecure, way of fixing the protocol that can still retain the efficiency of the original
protocol. Note that fully homomorphic
sorting is currently impractical to be deployed on a large scale \cite{HongKCLC21}.

Once the LBS receives $\llbracket z \rrbracket$, it can decrypt it to obtain $z$, and then subtract $2^l$ from $z$
to obtain the signed difference $d_i - d_j$, $1 \le i < j \le n$. This can be repeated for every pair of distances. We then end up
with $n(n-1)/2$ equations in the $n$ many $d_i$ $(1 \le i \le n)$. Hence, LBS will be able to solve for all the $d_i$ from this
overdetermined system of linear equations.

Once, say, $d_1$, is obtained. Then, the LBS can try to solve for $(X_{va},Y_{va})$ from the following equation:
\[
(X_{va} - x_1)^2 + (Y_{va} - y_1)^2 = d_1. 
\]
The above equation corresponds to a circle and there can be infinitely many solutions. Note that LBS knows
$(x_1,y_1)$, i.e., as plaintexts.
If $(X_{va},Y_{va})$ are encoded as (scaled) integers, then it will only have a ``couple'' of  solutions on an average \cite{KumaraswamyMV21}. But to
keep things simple, we can write similar equations for $d_2$, $d_3$, .... Since three circles are likely to intersect at
a single point, the LBS will very likely be able to recover the user virtual location $(X_{va},Y_{va})$. If there are more than 
two points at which these three circles intersect, then we can continue this process until we narrow down to a single point.

Once the LBS obtains $(X_{va},Y_{va})$, then it will try to recover the original user location coordinates $(X_{a},Y_{a})$. 
It is not unreasonable to assume that the LBS would have tried to recover the user location coordinates from the very
beginning. In this case, the LBS would also know the historical location coordinates $(X_{\eta-1},Y_{\eta-1})$, $\ldots$ ,
$(X_{\eta-t+1},Y_{\eta-t+1})$ used in Equations \eqref{eq:xva} and \eqref{eq:yva}. Also, the value of $t$ is typically
known to LBS as part of the protocol, or else, it can be guessed as it is usually small. Then, from 
Equations \eqref{eq:xva} and \eqref{eq:yva},
\[
X_a = t \cdot X_{va} - ( X_{\eta-1}  +  X_{\eta-2}  + \ldots +  X_{\eta-t+1}),
\]
\[
Y_a = t \cdot Y_{va} - ( Y_{\eta-1}  +  Y_{\eta-2}  + \ldots +  Y_{\eta-t+1}).
\]
Here, we are assuming that the virtual location information is only computed with the actual location data.
Else, what if initially the parameters $(X_{\eta-1},Y_{\eta-1})$, $\ldots$ , $(X_{\eta-t+1},Y_{\eta-t+1})$, were randomly chosen?
After $t$ instances of the protocol have been
evoked, the initially chosen random values will no longer affect the computation of $(X_{va},Y_{va})$. 
While the convergence and divergence of these moving averages is well-studied in statistics, we do not 
know of how to recover the individual data points, if at all it is possible. In this case, we can only recover
the virtual location coordinates.

\subsection {Another Failed Attempt}

Next, we look at another straightforward method to fix the comparison protocol of \cite{ma-infocom19}. 
The (signed) difference of distance is now masked by a random and independently chosen value $R$. Note that
 $R$ could be a possibly large value chosen independently for every difference. We then have  
\begin{equation} \label{eq:z-enc-rand}
\llbracket z \rrbracket = (\llbracket d_a \rrbracket - \llbracket d_b \rrbracket) \cdot \llbracket R \rrbracket.
\end{equation}
One would expect that the LBS upon decrypting $\llbracket z \rrbracket$ obtains
\begin{equation} \label{eq:z-rand}
 z  =  (d_a  -  d_b ) \cdot R, 
\end{equation}
and that this would only reveal the sign of $d_a  -  d_b$ and not the exact value, thereby, thwarting the attack mentioned previously.

We next show that the above method is insecure too. We use the technique from \cite{MurthyV19} to recover the difference
$d_a  -  d_b$ from $z$ alone with a good probability. The idea is to use the fact that every $d_a$ and $d_b$, $0 \le |d_a  -  d_b| \le m$. Therefore,
there will be a ``small'' factor of $z$ that is less than $m$ and, hence, feasible to recover this factor. This factor
would be a possible candidate for $R$. One could use the
brute-force technique to factorize, or, for larger values of $m$, the elliptic-curve method of factorization would be more efficient. Once a possible 
value of $R$ is determined, then 
\[
d_a  -  d_b = z/R.
\]
In case one ends up
with many candidates for $R$, then we need to brute force over these choices of $R$, and then check the consistency of these computed
differences with other similarly computed distances. This way inconsistent choices of $R$ are eliminated. In the worst case, we may end up with more than one possibility for the distances
$d_i$, and, hence, as many possibilities for $(X_{va},Y_{va})$. 

Hence this fix too would \textit{not} lead to a secure protocol.



\section {Conclusion}
In this paper, we analyzed the correctness of the protocol of Ma et al. \cite{ma-infocom19}. We showed that their
efficient ``secure'' comparison technique does not give the correct output. We then showed that straightforward attempts to fix this
flaw would lead to security vulnerabilities, where the location-based service provider will be able to recover
information about users locations. It seems that fixing the protocol of Ma et al. is non-trivial without 
incurring significant cost in terms of computation time and communication bandwidth.

There have been several attempts to design a privacy-friendly comparison protocol that is significantly more efficient than 
the homomorphic/MPC evaluation of the entire comparison circuit. Unfortunately, many of them have been shown to be insecure.
Hence, it is an interesting open problem to design a comparison protocol that is lightweight in terms of both time
and bandwidth.


\begin{acks}
This work was partially funded by the Infosys Foundation Career Development Chair Professorship grant for Srinivas Vivek.
\end{acks}

\bibliographystyle{ACM-Reference-Format} 
\bibliography{ref}

\end{document}